\documentclass[draft,jgrga]{agutex}

  \usepackage[dvipdf]{graphicx}

  \setkeys{Gin}{draft=false}
\authorrunninghead{NIKOLAEVA ET AL.}

\titlerunninghead{GERERATION OF STORM AND DRIVER TYPE}

\begin{document}

%
%

\title{Does generation of magnetic storm depend on type of solar wind?}
%
%

%
%




\authors{N. S. Nikolaeva, \altaffilmark{1}
Yu. I. Yermolaev, \altaffilmark{1} 
I. G. Lodkina , \altaffilmark{1} 
}

\altaffiltext{1}{Space Plasma Physics Department, Space Research Institute, 
Russian Academy of Sciences, Profsoyuznaya 84/32, Moscow 117997, Russia. 
(nnikolae@iki.rssi.ru)}






%
%


\begin{abstract} 

The purpose of this work is to draw attention of readers to a problem of 
possible differences in generation of magnetic storms induced by various 
large--scale solar wind (SW) streams: CIR, Sheath and ICME (including MC 
and Ejecta). Recently we showed that when using a modification of 
formula by \citet{Burtonetal1975} for connection of interplanetary 
conditions with $Dst$ and $Dst^*$ indices the efficiency of storm generation 
by Sheath and CIR is $\sim 50\%$ higher than generation by ICME 
\citep{Nikolaevaetal2013,Nikolaevaetal2015}. 
In the literature there are many various functions 
coupling (FC) various interplanetary parameters with magnetospheric 
state. In this work we study the efficiency of main phase storm 
generation by different SW streams when using 12 another FCs on the 
basis of OMNI data during 1976--2000. Obtained results show that for most 
part of FCs Sheaths have the highest efficiency and MCs have the lowest 
efficiency in accordance with our previous results. The reliability of 
the obtained data and possible reasons of divergences for various FCs 
and various SW types require further researches.

\end{abstract}

%
%

%

\begin{article}

%
%

\section{Introduction}

One of the unsolved problems of magnetosphere is its reaction on the solar wind (SW) variations.
On the one hand, in the literature there are many functions coupling (FC) interplanetary 
conditions with magnetospheric state. 
The most of them are different functional forms of the solar wind electric 
field $Ey$, expressed through various parameters of the SW stream 
\citep{Burtonetal1975,KanLee1979,Hardyetal1981,HolzerSlavin1982,Wygantetal1983,Newelletal2007,BorovskyBirn2014}.
Another approach was used by 
\citet{Borovsky2008,Borovsky2013a,Borovsky2013b,Borovsky2014}: 
they estimated how local plasma parameters 
near the reconnection region (between magnetosheath and magnetosphere) control the 
reconnection on the dayside of the magnetosphere. 
He found the formula for local reconnection rate at the dayside of the magnetosphere 
$Rquick$ by using the equation of \citet{CassakShay2007} for local reconnection 
rate between the two asymmetric plasmas near the magnetopause boundary. 
Additionally it was obtained the coupling function $FC = Vsw + 56 Bs$, which does not have 
clear physical interpretation, but gives the more higher correlation with geomagnetic 
indices than any of electric field functions or reconnection functions \citep{Borovsky2014}.
These FCs are usually used for all available data during long intervals of measurements.

On the other hand, one of the recent experimental facts is that magnetic storms 
generated by different types of solar wind streams are different 
\citep{BorovskyDenton2006,Huttunenetal2006,Pulkkinenetal2007,Yermolaevetal2007,PlotnikovBarkova2007,Longdenetal2008,Turneretal2009,Guoetal2011,Nikolaevaetal2013,Nikolaevaetal2014,Nikolaevaetal2015,Yermolaevetal2010,Yermolaevetal2014,Yermolaevetal2015}.
In particular, it was shown that coefficients $CE$ (and $CE^*$) of linear relation between
$Dst$ (and $Dst^*$) index and integral of interplanetary electric field $Ey = Vx Bz$ 
\citep{Nikolaevaetal2013,Nikolaevaetal2015} depend on type of solar wind stream and 
at the same integral of $Ey$ Sheath and CIR generate $\sim 1.5$ stronger magnetic storms 
than MC and Ejecta (see Table 1).

In discussed papers \citep{Nikolaevaetal2013,Nikolaevaetal2015} a modification of function by 
\citet{Burtonetal1975} connected $Dst$ and $Dst^*$ indices and interplanetary electric field 
$Ey$ was used.
Other FCs mentioned above have not been used for analysis of development of magnetic storms
induced by various SW streams and for comparison of efficiencies of storm generation by 
these SW drivers. 
The aim of this brief paper is to check the dependence of efficiency of magnetic storm generation 
on type of solar wind stream using other coupling functions. 

\section{Data and methods} 

For analysis we use SW and IMF parameters of OMNI dataset  
(http://omniweb.gsfc.nasa.gov) \citep{KingPapitashvili2004} and Kyoto dataset of $Dst$ index 
measurements
(http://wdc.kugi.kyoto-u.ac.jp/index.html). On the basis of these data we prepared
the Catalog of large-scale interplanetary events for period of 1976--2000 and 
we select main phases of magnetic storms ($Dst_{min} \le -50$ nT), 
generated by MC (10 storms, 77 1-h points), Ejecta (31 storms, 324 1-h points), 
Sheath (21 storms, 166 1-h points), CIR (31 storms, 279 1-h points) 
\citep{Yermolaevetal2009,Nikolaevaetal2013,Nikolaevaetal2014,Nikolaevaetal2015}. 

The full set of 12 coupling functions is presented in the first column 
of the Table 2. 
They include 8 coupling functions (with our numbers FC1 -- FC7, FC9) that were taken 
from paper \citep{BorovskyBirn2014},
and 4 coupling functions (FC8, FC10, FC11, FC12) were suggested by other authors:

The function FC1 = $\sin^{2}(\theta/2)$ is the pure geometrical one; here clock angle $\theta$ is 
the IMF polar angle projected on $(Y-Z)_{GSM}$ plane (for example, 
 \citep{Sonnerup1974,Newelletal2007,Wilderetal2011,BorovskyBirn2014}).   
Three coupling functions  FC2 = $vByz$, FC3 = $vBz$, FC4 = $vBs$ are the different variants 
of electric field presentation by transverse component of IMF $Byz = (By^2 + Bz^2)^{1/2}$, 
by $Bz$ component of IMF, and by $Bs$ southward component of IMF
\citep{Burtonetal1975,Hardyetal1981,HolzerSlavin1982}.
Two coupling functions FC5 = $vByz \sin^{2}(\theta/2)$ and FC6 = $vByz \sin^{4}(\theta/2)$ 
are the electric field presentations with including clock angle dependence 
\citep{KanLee1979,Wygantetal1983}.
The coupling function FC7 = $v^{4/3} Byz^{2/3} \sin^{8/3}(\theta/2)$ is 
the rate magnetic flux opened at the magnetopause \citep{Newelletal2007}. 
It has the best correlation with nine from ten indices
(with exception of Dst index) and present a nearly universal SW--Magnetosphere 
coupling function obtained from 
ten magnetospheric state variables \citep{Newelletal2007}.  
The coupling function FC9 = $Rquick$ 
is the reconnection rate on the dayside of 
the magnetosphere obtained with methodology of Borovsky (see, paper
\citep{BorovskyBirn2014} and references inside). 
$Rquick \sim \sin^{2}(\theta/2) C^{-1/2} n^{1/2} v^2 (1+ \beta_{s})^{-3/4}$, where
$\beta_{s}$ is the plasma beta of magnetosheath, 
$C$ is the compression ratio of the bow shock.
Both $\beta_{s}$ and $C$ are functions of  the Alfven Mach number $M_{A}$
(see, expressions (3)--(8) in paper \citep{BorovskyBirn2014}).
FC8 = $p^{1/2} v^{4/3} Byz^{2/3} \sin^{8/3}(\theta/2)$ is the best coupling function
for $Dst$ index prediction which differs from nearly universal FC7 
inferred from 10 magnetospheric 
state variables by dynamic pressure factor $p^{1/2}$ 
(correction on the magnetopause currents) \citep{Newelletal2007}. 
FC10 = $p^{1/2} v^{4/3} Byz \sin^{6}(\theta/2)$  is the best function for $Dst$ index prediction 
in non-linear dynamic systems 
\citep{TemerinLi2006,Balikhinetal2010} 
and includes the correction on dynamic pressure and the IMF clock-angle $\theta$ dependence 
as $ \sin^{6}(\theta/2)$; it was followed from model by \citet{KanLee1979} when dependence 
of reconnection line length on clock angle $\theta$ was included \citep{Balikhinetal2010}.
FC11 = $Vsw + 56 Bs$ is the function without any clear physical interpretation, or mathematical 
variant with the best correlation \citep{Borovsky2014}. 
$FC12 = M_{A}$ is the Alfven Mach number (the ratio of SW velocity
to Alfven speed). 
The dependence on $M_{A}$ is included in function FC9 = $Rquick$ 
(see, \citep{BorovskyBirn2014}), but FC12 allows one to analyse
only dependence on $M_{A}$. 

To compare the interplanetary drivers we estimate an efficiency of magnetic storm generation 
by type of solar wind stream with using 12 coupling functions. For each type of the SW stream 
the coefficients, $C_{FCN}$ and $C_{FCN}^*$ were obtained by the linear approximation 
of $Dst$ and $Dst^*$ indices: 
$Dst_{i} = C_{0} + C_{FCN} FCN_{i}$,  
$Dst^*_{i} = C_{0}^* + C_{FCN}^* FCN_{i}$  
where $i$ is the point number at the main phase of magnetic storms 
induced by given type of SW stream,
and FCN are coupling functions FC1 -- FC12.
Obtained approximation coefficients $C_{FCN}$ and $C_{FCN}^*$
are interpreted as efficiencies 
of $Dst$ and $Dst^*$ generation and presented in Table 3. 
The correlation coefficients 
for these fits which are connected with the accuracy of the description of  
SW-magnetosphere connections for corresponding coupling function, 
change in rather wide interval - from 0.01 up to 0.63 (see Figure 1). As mentioned above, 
the statistics of points for approximation is rather high - from 77 points 
for MC up to 329 for Ejecta (see, Table 3). So, in most cases of calculated coefficients 
$C_{FCN}$ and $C_{FCN}^*$ are obtained with good statistical significances. 
Reliability of coefficients $C_{FCN}$ and $C_{FCN}^*$ are connected with many 
parameters (data series, statistics, type of coupling function etc.) and requires 
further investigations which will be made in our future works. 
To obtain a rough estimation of a total (on all 12 coupling functions)  
characteristics of efficiency of magnetic storm generation by different types of solar 
wind stream, the absolute values of approximation coefficients for each 
coupling function were ranked on types of the solar wind stream, i.e. 
values from 1 (for the lowest coefficient) to 4 (for the highest 
coefficient) were appropriated to the corresponding efficiencies (see numbers 
in parentheses). Last line of Table 3 presents the average values of efficiency defined by 
this method which maybe called "place number" method. Because the average value 
of sequence 1, 2, 3, and 4 is 2.5 ((1+2+3+4)/4=2.5), the average values over $12 FCs >2.5$ means 
that for given SW type the most part of FCs have high efficiencies. 

\section{Results} 

Table 2 presents the average values (signed by $<>$), standard deviations and medians of 
values of twelve coupling functions FC1 -- FC12 calculated at the main phase of magnetic 
storms induced by 4 types of SW drivers. 
These data show that all FCs have large deviations of values and average values 
can depend on SW type.
Behaviour of FCs at the main phase of storms is determined by changes of corresponding 
SW and IMF parameters (see, for instance, average dynamics 
of parameters for CIR, Sheath, MC and Ejecta in papers by 
\citet{Yermolaevetal2010,Yermolaevetal2015}).

The Table 3 data allow one to estimate the efficiency of the magnetic storm generation 
by the different SW types. The lowest negative coefficients values $C_{FCN}$ and 
$C_{FCN}^*$ (or the highest values of their magnitude) indicate the highest efficiencies 
of SW type for given coupling function FCN, because at the equal values of FCN the calculated 
value of $Dst$ (or $Dst^*$) is lower (magnetic storm is stronger). 
In the contrast with the most values of $C_{FCN}$ and $C_{FCN}^*$ magnitude, 
the lowest values of them indicate the lower values of efficiencies. 
Table 3 shows that for each coupling function the different SW types can be the most 
effective for development of main phase of magnetic storm. 

For description of $Dst$ at the main phase of Sheath-driven storms
the most efficiency is associated with five FCs (FC4, FC6, FC7, FC9, FC11); 
for CIR-driven storms - two FCs (FC2, FC8); for the MC-driven storms - two FCs 
(FC3, FC12), and for Ejecta-driven storms - three FCs (FC1, FC5, FC10). 
Similarly for corrected (on dynamic pressure) $Dst^*$ at the main phase of Sheath-driven 
storms the most efficiency is related with seven FCs 
(FC3, FC4, FC5, FC6, FC7, FC9, FC11);
for CIR-driven storms - only FC2; for MC-driven storms - only FC12, for Ejecta-driven 
storms - two FCs (FC1, FC8).
In particular, we can see in Table 3 that the coefficients of linear relation between 
$Dst$ (and $Dst^*$) indices and coupling function FÑ9 = $Rquick$ depend on type of solar 
wind stream, and Sheath and CIR generate $\sim$ 1.42 and $\sim$ 1.38 stronger magnetic 
storms than MC in agreement with data presented in Table 1 
\citep{Nikolaevaetal2013,Nikolaevaetal2015}. 

Because of different dimensionality of  $C_{FCN}$ and $C_{FCN}^*$, their different ranges of 
changes, different distributions and so on, it is impossible to directly compare them for 
different FCs. 
To overcome these difficulties we use method of "place number" described in the Data and 
Method Section. 
This method allows us to obtain 
that in average Sheath have more large efficiency of the magnetic storm generation 
and MC have more low efficiency (in factors $\sim$ 1.4 and $\sim$ 1.7 for $Dst$  
and $Dst^*$, respectively) in agreement with our prevous results. 

\section{Discussion} 

The carried-out analysis allowed us to obtain rather unexpected results. 
On the one hand, the found distinctions in efficiency of storm 
generation with various drivers allow to answer the question raised in 
the article title positively: Yes, generation of magnetic storms depends 
on the type of solar wind. On the other hand, ratios of efficiencies of 
various SW types differ for various functions (or physical models) 
connecting parameters of the interplanetary space and calculated 
(modeled) values of Dst and Dst* indices. The answer to a question "Why 
it so happens?" is beyond the scope work. Here we will discuss only the 
general approaches to the solution of this problem.

It should be noted that the dependence of efficiency of magnetic storm 
generation by different  SW types on type of coupling function can be 
associated with following causes:

1.  An accuracy of physical process description.
Different coupling functions (or used physical models) have different 
accuracy of geomagnetic index prediction on the basis of measured 
interplanetary plasma and IMF parameters.

2. Used data set.
Obtained results are related with specific data selection. The stability 
of obtained results with other data sets requires further study.

3. An accuracy of data approximation.
Although the approximation accuracy is directly related with two 
previous problems, its analysis could provide additional information for 
a more reliable conclusion about the differences between efficiency of 
magnetic storms generation by different SW types.

The correlation coefficients are one of the criterion of the accuracy of data 
approximation and correlation coefficients between measured $Dst$ and pressure corrected  
$Dst^*$ indices and 12 coupling functions during main phases of magnetic storms driven by 
different solar wind streams are presented in Figure 1. 
The most part of correlation coefficients has sufficiently high values. In particular  
the highest values of coefficients ($\sim 0.5$ up to $\sim 0.63$) are observed 
for Sheath- driven storms. 
So, conclusion about high efficiency of Sheath is reliable. In a small part of low
coefficients it is necessary to increase the number of magnetic storms to increase
the statistical significance of results.
Nevertheless, we believe that obtained results can be considered as a basis for further
investigation. 

\section{Conclusions} 

Thus on the OMNI data for interval 1976--2000 we study the generation of magnetic 
storms induced by MC (10 storms, 77 points), Sheath (21 storms, 166 points), 
Ejecta (31 storms, 324 points), CIR (31 storms, 279 points) and
evaluate the dependence of efficiency of magnetic storm generation on type of 
SW stream using 12 coupling functions.
Also we estimated the correlation coefficients between 12 coupling functions 
and measured $Dst$ and pressure corrected $Dst^*$ indices 
during the main phases of magnetic storms.

Our study allowed to obtain the following results 

1) The generation of magnetic storm depends on type of solar wind 
in agreement with previous results 
\citep{BorovskyDenton2006,Huttunenetal2006,Pulkkinenetal2007,Yermolaevetal2007,PlotnikovBarkova2007,Longdenetal2008,Turneretal2009,Guoetal2011,Nikolaevaetal2013,Nikolaevaetal2014,Nikolaevaetal2015,Yermolaevetal2010,Yermolaevetal2014,Yermolaevetal2015}.

2) The most part of the coupling functions have the high efficiency for 
Sheath in agreement with our result presented in previous papers 
\citep{Nikolaevaetal2013,Nikolaevaetal2015}. 

3) In contrast with Sheath the most part of coupling functions for MC have
the lowest efficiencies. This also confirm our results 
\citep{Nikolaevaetal2013,Nikolaevaetal2015}. 

4) Efficiencies of magnetic storm generation by CIR and Ejecta are intermediate. 

We consider that the results presented here are preliminary: 
the confirmation of these results, increase in their reliability require 
further investigation, and we plan to continue this analysis.


%
%
%
%
%
%
%

\begin{acknowledgments}
The authors are grateful for the opportunity to use the OMNI database. The OMNI data were 
obtained from GSFC/ SPDF OMNIWeb (http://omniweb.gsfc.nasa.gov). This work was supported 
by the Russian Foundation for Basic Research, projects 13--02--00158 and 16--02--00125, and by 
Program of Presidium of the Russian Academy of Sciences. 
\end{acknowledgments}

\end{article}


%
%

%
%
%
%
%


\begin{figure}
\noindent\includegraphics[width=10cm]{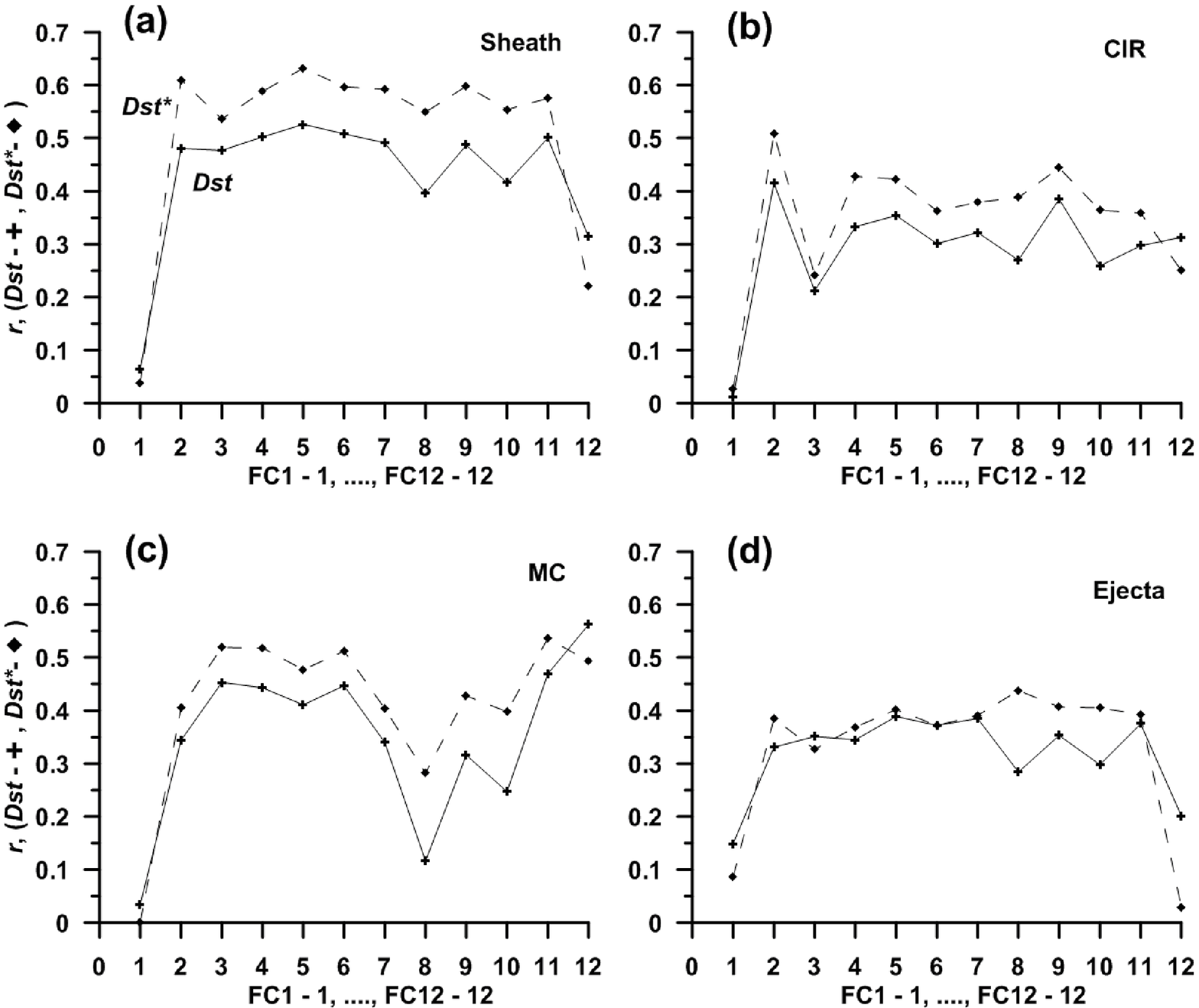}
\caption{Correlation coefficients, $r$, between measured $Dst$ (crosses, and solid line) and pressure corrected $Dst^*$ (diamonds, and dotted line) indices and the values of 
12 coupling functions FC1--FC12 during the main phases of magnetic storms driven by: (a) Sheath; (b) CIR; (c) MC; (d) Ejecta}
\end{figure}

%

\begin{table}
{\small
\caption{The coefficients $CE$ (and $CE^*$) of linear relation between $Dst$ and $Dst^*$ indices and integral of interplanetary electric field $Ey = Vx Bz$ at the main phases of magnetic storms driven by different solar wind streams \citep{Nikolaevaetal2013,Nikolaevaetal2015}.}
\centering
\begin{tabular}{c|c|c|c|c}
\hline
    SW type  &      MC       &      Ejecta  &   Sheath &  CIR  \\
\hline
  $CE$ (for $Dst$)  & $–2.55 \pm 0.75$ &  $–2.3 \pm 1.0$ &  $–3.2 \pm 1.6$  & $–2.8 \pm 1.1$ \\
\hline
  $CE^*$ (for $Dst^*$) & $–2.0 \pm 1.1$   & $–2.1 \pm 1.1$  & $–3.4 \pm 1.9$  & $–3.0 \pm 1.5$ \\
\hline
\end{tabular}
}
\end{table}

\begin{table}
{\footnotesize
\caption{The average values of 12 coupling functions with standard deviation and median values 
at the main phases of magnetic storms driven by MC, Ejecta, Sheath, CIR.
}
\centering
\begin{tabular}{l|cc|cc|cc|cc}
\hline
FC    &  \multicolumn{2}{c}{MC-storms} & 
               \multicolumn{2}{|c}{Ejecta-storms}  &
               \multicolumn{2}{|c}{Sheath-storms}  &
	\multicolumn{2}{|c}{CIR-storms}  \\
type      &  $<>$  & median & $<>$ & median & $<>$ & median & $<>$ & median \\
\hline
FC1=$\sin^{2}(\theta/2)$    & 0.91 $\pm$ 0.11 & 0.93 & 0.84 $\pm$ 0.19 & 0.92 & 0.82 $\pm$ 0.19 & 0.87 & 0.795 $\pm$ 0.22 & 0.88  \\
FC2=$vByz$    & 4.61 $\pm$ 2.39 & 3.86 & 3.83 $\pm$ 1.49 & 3.69 & 5.66 $\pm$ 3.86 & 4.40 & 4.29 $\pm$ 1.97 &  3.91  \\
FC3=$v Bz$     & 3.69 $\pm$ 1.87 & 3.26 & 2.60 $\pm$ 1.80 & 2.60 & 3.50 $\pm$ 3.36 & 3.05 & 2.56 $\pm$ 2.36 & 2.62  \\
FC4=$v Bs$     & 3.74 $\pm$ 1.82 & 3.23 & 2.87 $\pm$ 1.48 & 2.66 & 3.93 $\pm$ 2.9 & 3.28 & 3.16 $\pm$ 1.87 & 2.81  \\
FC5= $v Byz \sin^{2}(\theta/2)$  & 4.15 $\pm$ 2.04 & 3.44 & 3.21 $\pm$ 1.45 & 3.06 & 4.58 $\pm$ 3.29 & 3.59 & 3.42 $\pm$ 1.87 & 3.05  \\
FC6= $v Byz \sin^{4}(\theta/2)$  & 3.78 $\pm$ 1.87 & 3.26 & 2.83 $\pm$ 1.49 & 2.65 & 3.91 $\pm$ 3.03 & 3.19 & 2.94 $\pm$ 1.91 & 2.64  \\
FC7= $v^{4/3} Byz^{2/3} \sin^{8/3}(\theta/2)$  & 13.83 $\pm$ 5.30 & 12.35 & 11.70 $\pm$ 4.95 & 11.04 & 14.86 $\pm$ 9.35 & 12.81 & 11.49 $\pm$ 5.61 & 10.92 \\
FC8=$p^{1/2}v^{4/3}Byz^{2/3}\sin^{8/3}(\theta/2)$ & 25.02 $\pm$ 12.09 & 21.36 & 23.69 $\pm$ 13.74 & 21.28 & 37.01 $\pm$ 40.68 & 23.91 & 26.6 $\pm$ 16.3 & 21.83 \\                         
FC9= $Rquick$    & 172.43 $\pm$ 69.76 & 149 & 147.4 $\pm$ 58.4 & 136 & 194.83 $\pm$ 122.94 & 158 & 159.4 $\pm$ 79.35 & 142.36 \\
FC10= $p^{1/2} v^{4/3} Byz \sin^{6}(\theta/2)$  & 47.74 $\pm$ 28.34 & 38.89 & 40.41 $\pm$ 30.85 & 32.69 & 68.52 $\pm$ 84,51 & 40.30 & 46.92 $\pm$ 40.52 & 32.84 \\
FC11= $Vsw + 56 Bs$  & 910.3 $\pm$ 231.2 & 850 & 792.7 $\pm$ 198.01 & 780 & 904.1 $\pm$ 349.2 & 850 & 801.5 $\pm$ 251.34 & 769 \\
FC12= $M_{A}$  & 6.90 $\pm$ 3.36 & 6.3 & 8.45 $\pm$ 3.38 & 7.7 & 7.42 $\pm$ 3.59 & 6.5 & 8.70 $\pm$ 3.85 & 7.66 \\
\hline
\end{tabular}
}
\end{table}

\begin{table}
{\footnotesize
\caption{Coefficients of the linear relation between the $Dst$ and $Dst^*$ indices and the 
coupling function values at the main phase of magnetic storms driven by 4 types of the SW.
}
\centering
\begin{tabular}{l|cc|cc|cc|cc}
\hline
FC      &  \multicolumn{2}{c}{MC, 77 points} & 
               \multicolumn{2}{|c}{Ejecta, 324 points}  &
	\multicolumn{2}{|c}{Sheath, 166 points} &
                \multicolumn{2}{|c}{CIR, 279 points}     \\                            
type      &  $C_{FCN}$  & $C_{FCN}^*$ & $C_{FCN}$  & $C_{FCN}^*$ & $C_{FCN}$  & $C_{FCN}^*$ & $C_{FCN}$  & $C_{FCN}^*$  \\
\hline

FC1  & -8.01 (2) &  0.26 (1) &  -20.6 (4) &  -12.8 (4) & -14.4 (3) &  -9.11 (3) & -1.79 (1) &  4.58 (2) \\
FC2  & -3.77 (1) & -4.41 (1) & -5.88 (3)  &  -7.35 (2) &  -5.35 (2) & -7.38 (3) & -7.16 (4)  & -9.20 (4) \\
FC3  & -6.35 (4) & -7.23 (3) & -5.19 (2) &  -5.18 (2) &  -6.09 (3)  & -7.44 (4) & -3.03 (1) & -3.75 (1) \\
FC4  & -6.37 (4) & -7.42 (2) & -6.23 (2)  &  -7.18 (1) & -7.20 (4)  & -9.25 (4) & -5.81 (1) & -8.05 (3) \\
FC5  & -5.29 (1) & -6.10 (1) &  -7.13 (4) &  -7.91 (2) & -6.86 (3)  & -8.96 (4) & -6.42 (2) & -8.19 (3) \\
FC6  & -6.26 (2) & -7.14 (3) &  -6.65 (3) &  -7.13 (2) & -7.17 (4)  & -9.16 (4) & -5.36 (1) & -6.92 (1) \\
FC7  & -1.68 (1) & -1.99 (1) & -2.06 (3)  & -2.25 (2) & -2.25 (4) & -2.95 (4) & -1.95 (2) & -2.46 (3)  \\
FC8   & -0.25 (1) & -0.61 (1) & -0.55 (3) & -0.91 (4) & -0.42 (2) & -0.63 (2) & -0.56 (4) & -0.85 (3) \\
FC9   & -0.12 (1) & -0.16 (1) & -0.16 (2) & -0.20 (2) & -0.17 (4) & -0.23 (4) & -0.16 (3) & -0.20 (3)  \\
FC10  & -0.23 (3) & -0.36 (3) & -0.26 (4) & -0.37 (4) & -0.21 (1) & -0.31 (1) & -0.22 (2) & -0.32 (2) \\
FC11  & -0.05 (3) & -0.06 (3) & -0.05 (2) & -0.06 (2) & -0.06 (4) & -0.08 (4) & -0.04 (1) & -0.05 (1) \\
FC12  & 4.38 (4)  &  3.86 (4) &  1.59 (1)  &  0.25 (1) &  3.75 (3) &  2.86 (3) &  2.76 (2) & 2.28 (2) \\                
\hline
rel. aver    &   2.17  &  2.00  &  2.75  &  2.33  &  3.08  &  3.33  &  2.00  &  2.33 \\
\hline
\end{tabular}
}
\end{table}

\end{document}